\documentclass[trackchanges, twocolumn]{aastex701}

\usepackage{placeins}
\usepackage{amsmath}

\begin{document}

\title{Probabilistic Inference of Cosmological Density Parameters from Synthetic Hubble Expansion Data of Varying SNR Using Diverse Artificial Neural Network Architectures}

\author[orcid=0009-0003-2374-5363]{Zijian Jin}
\affiliation{Southridge School, 2656 160 St, Surrey, BC V3Z0B7, Canada}
\email[show]{ajin@southridge.ca}

\author[orcid=0000-0001-9214-7437]{Jaehyon Rhee} 
\affiliation{Center for Astrophysics $\vert$ Harvard \& Smithsonian, 60 Garden Street, Cambridge, MA 02138, USA}
\email[show]{jrhee@cfa.harvard.edu}

\begin{abstract}

This paper builds upon ParamANN's novel approach \citep{Pal2024} of using ANNs to infer cosmological density parameters by determining optimal architecture for varying synthetic Hubble data SNRs in estimating the density parameters $\Omega_{m, 0}$ and $\Omega_{\Lambda, 0}$ across redshift values $z \in [0, 1]$. To generate the synthetic data, this study randomly sampled initial free parameter values at $z=0$ from theoretically motivated priors and evolved them backwards using the first Friedmann Equation to generate clean $H(z)$ curves. Then, this paper adds realistic noise of high, normal, and low SNR by sampling relative uncertainties from a Gaussian KDE on 47 real data observations compiled by \citet{Bouali2023}. In the end, this study found that a RNN that uses BiLSTM is the most effective for high and normal SNR data across four quantitative metrics. On the other hand, a combination of convolution and recurrent layers that uses GRU performed the best for low SNR data across the same four metrics. A comparison between the results of this paper’s ANN predictions and those of ParamANN shows that all architectures tested in this paper regardless of training SNR are statistically consistent within 1 standard deviation of ParamANN. However, most ANN results are not statistically consistent within 3 standard deviations of \citet{Planck2020}, showing a significant difference between ANN and the more traditional MCMC methods used by Planck collaboration.

\end{abstract}

\section{Introduction} \label{sec:introduction}

The Hubble tension refers to the significant discrepancy between local measurements of the Hubble constant and those derived from the data on the Cosmic Microwave Background (CMB) from the early universe within the standard $\Lambda\texttt{CDM}$ model \citep{Planck2016, Riess2018, Riess2019, Planck2020, Riess2020, Wong2020, DiValentino2021, Riess2021, Riess2022}.  As the possibility of measurement technique flaws is ruled out, there is growing confidence that the standard $\Lambda\texttt{CDM}$ model, which defines cosmological density parameters like $\Omega_{m}$ and $\Omega_{\Lambda}$, may be incomplete or incorrect, suggesting new physics beyond it \citep{Riess2020, Vagnozzi2020, Hu2023}.  It is crucial to measure cosmological density parameters accurately because comparing them with theoretical predictions allows us to validate or refine the standard model of cosmology to predict the past and future of the universe and understand phenomena such as dark matter and dark energy.

Cosmological density parameters are dimensionless ratios comparing the different energy density components of the universe to the critical density required for a flat universe at $z=0$. Specifically, the five main density parameters of interest are $\Omega_{r, 0}$, $\Omega_{m, 0}$, $\Omega_{b, 0}$, $\Omega_{k, 0}$, and $\Omega_{\Lambda, 0}$, representing the relative proportion of the universe's energy density at $z=0$ as a result of radiation, matter, baryon matter, curvature, and dark energy, respectively.

Using the values of these density parameters, we can then calculate the Hubble constant, which measures the rate of the universe's expansion, at any given redshift using the first Friedmann equation:

\begin{equation}
\begin{aligned}
H(z) = H_0 \Big[ &
  \Omega_{r,0}(1+z)^4 + \Omega_{m,0}(1+z)^3 \\
& + \Omega_{k,0}(1+z)^2
  + \Omega_{\Lambda,0}(1+z)^{3(1+w)}   
\Big]^{1/2}.
\end{aligned}
\label{eq:friedmann-full}
\end{equation}

It is commonly agreed in past literature that $\Omega_{r, 0}$ and $\Omega_{k, 0}$ are very close to zero, meaning we will fix the values of those two density parameters to be 0 in this paper. Furthermore, we will also be fixing the value of $H_0$ to be $68$ $\text{km/s/Mpc}$, which is suggested by analyzing data from the CMB \citep{Planck2020, Freedman2023}. As a result, the only two free density parameters this study will estimate are $\Omega_{m, 0}$ and $\Omega_{\Lambda, 0}$. We see that a corollary from these assumptions is that $\Omega_{m, 0} + \Omega_{\Lambda, 0} = 1$. 

Furthermore, this paper will use the typically accepted $\Lambda\texttt{CDM}$ model of the universe, in which the universe is spatially flat with $w=-1$. Therefore, by plugging $\Omega_{r, 0} = \Omega_{k, 0} = 0$ and $w=-1$ into Eq.~\eqref{eq:friedmann-full}, we get

\begin{equation}\label{eq:friedmann-simplified}
H(z) = H_0 \sqrt{\Omega_{m,0}(1+z)^3 + \Omega_{\Lambda,0}}.
\end{equation}

\noindent where $H_0 = 68$ $\text{km/s/Mpc}$.

From Eq.~\eqref{eq:friedmann-simplified}, we see that we can easily map $(\Omega_{m, 0}, \Omega_{\Lambda, 0}, z)  \xrightarrow{\text{Eq.~\eqref{eq:friedmann-simplified}}}H(z)$ for the universe model assumed since there is a closed form formula. However, the inverse mapping $H(z) \xrightarrow{} (\Omega_{m, 0}, \Omega_{\Lambda, 0}, z)$ does not have a closed-form solution. This is problematic because we can currently only measure $H(z)$ with observational methods like baryon acoustic oscillations (BAO) \citep{Alam2017, Alam2021eBOSS}, Type Ia supernovae \citep{Brout2022PantheonPlusCosmo, Scolnic2022PantheonPlusData}, CMB \citep{Planck2020}, weak lensing \citep{Abbott2022}, etc., but not $\Omega_{m, 0}$ or $\Omega_{\Lambda, 0}$ directly, meaning finding the values of those density parameters is nontrivial. 

Past research used statistical methods like Markov Chain Monte Carlo (MCMC) to do this inverse-mapping. For instance,\citet{Planck2020}, the most comprehensive study on estimating the density parameter values to date, uses the Metropolis-Hastings algorithm, an MCMC method. However, despite MCMC's advantages, such as a lower statistical uncertainty, it is computationally expensive, sometimes imprecise, and is very sensitive to hyperparameters. As a result, cosmologists have tried using alternative methods for this task in recent years.

For example, \citet{Pal2024} used a novel approach of training an artificial neural network (ANN) called ParamANN to estimate the means and standard deviations of $\Omega_{m, 0}$, $\Omega_{\Lambda, 0}$, $\Omega_{k, 0}$, and $H_0$. Specifically, the architecture of ParamANN consists of an input layer, 1 hidden dense layer, and an output layer with 8 neurons (representing the means and the standard deviations of the four free parameters it tries to estimate). Unfortunately, one of the key challenges in enabling ANNs currently is the lack of real training data points. As a result, ParamANN generated additional synthetic data points by extrapolating the few real observations collected by \citet{Moresco2016} (through Cosmic Chronometers) using a covariance matrix. 

In this paper, we will generate synthetic data points by extrapolating 47 real Hubble expansion observations compiled by \citet{Bouali2023}, which includes more data points than \citet{Moresco2016}, as it also utilized BAO for finding $H(z)$ observations. Furthermore, this paper aims to improve upon ParamANN's novel approach by 1) testing more complex ANN architectures, such as those with convolutional and recurrent layers, to find the most optimal architecture based on quantitative evaluation metrics, 2) examining how the signal-to-noise ratio (SNR) of the synthetic training data affects each ANN architecture's performance, and 3) using more real data points to generate the synthetic training data. However, it is important to note that ParamANN outputs the predicted values for four cosmological density parameters. In contrast, this paper will only be focused on two density parameters ($\Omega_{m, 0}$ and $\Omega_{\Lambda, 0}$). Additionally, the uncertainty calibration method for each estimated value between this paper and ParamANN also differs because we used a different set of real observations to generate synthetic data.

\section{Methodology}

\subsection{Noise Model}

A major problem with the ANN approach is that there has only been 47 real data points of $H(z)$ for $0 \le z \le 1$ compiled by \citet{Bouali2023} from \citet{ChimentoForte2008}, \citet{Gaztanaga2009}, \citet{Stern2010},  \citet{Blake2012}, \citet{Moresco2012}, \citet{Chuang2013}, \citet{ChuangWang2013}, \citet{FontRibera2014}, \citet{Oka2014}, \citet{Zhang2014}, \citet{Delubac2015}, \citet{Moresco2016}, \citet{Alam2017}, \citet{Bautista2017}, \citet{Ratsimbazafy2017}, and \citet{Wang2017}. As mentioned in Section~\ref{sec:introduction}, this is not enough real data to train a neural network model. Therefore, to address this problem, this paper uses statistical methods to generate synthetic Hubble expansion data with realistic noise added, giving the ANN models more training data.

\begin{figure}
  \plotone{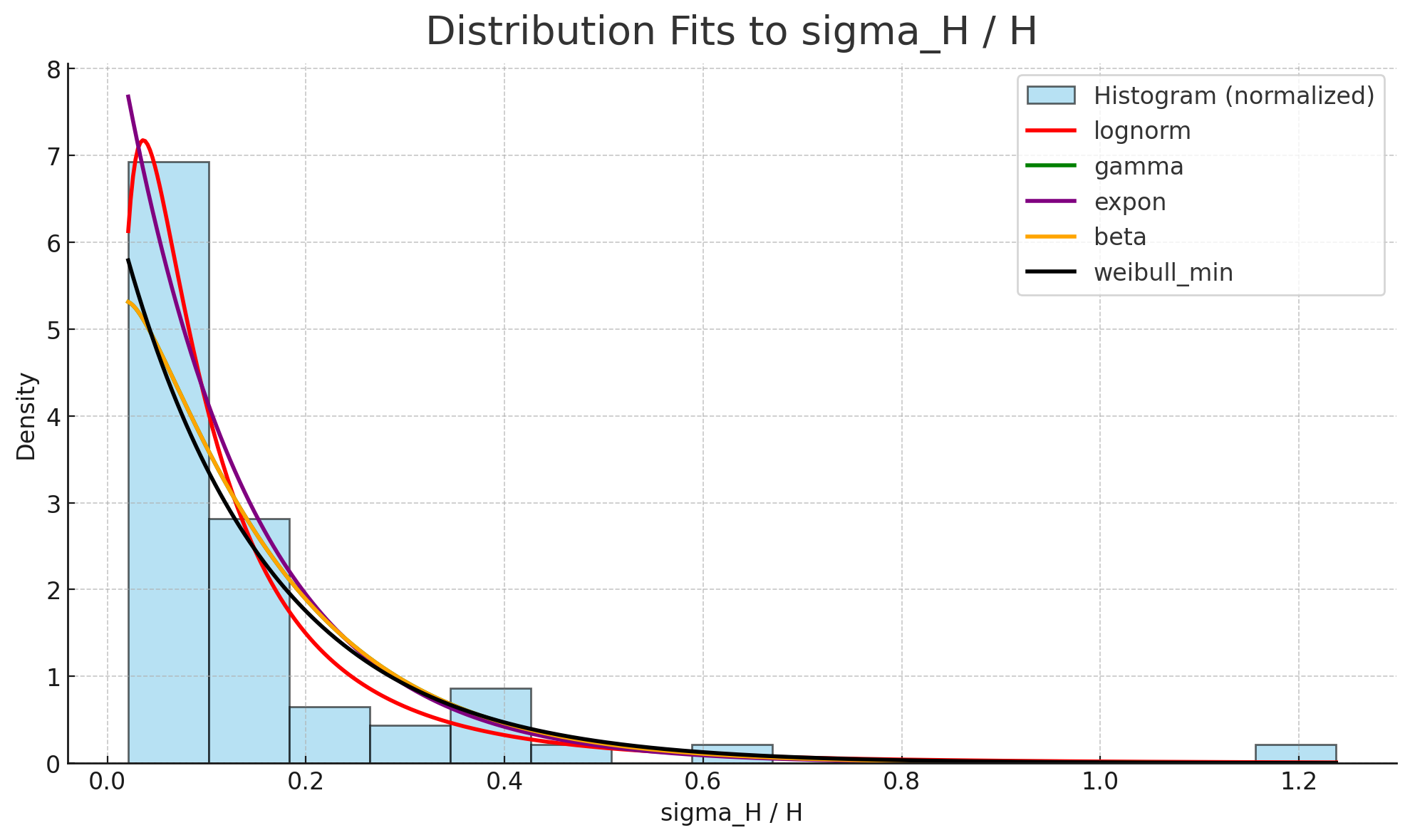}
  \caption{
  \label{fig:uncertainty-distribution-fits}
  Density histogram of the relative uncertainty of the real data compared to common distribution fits.}
\end{figure}

Fortunately, the real data provided by \citet{Bouali2023} includes the absolute uncertainty $\sigma_H(z)$ for the Hubble constant at each point $z$. From here, we can add the relative uncertainty for each $z$ by simply noting that
\begin{equation}\label{eq:sigrel}
\sigma_{H,\mathrm{rel}}(z)=\frac{\sigma_H(z)}{H(z)}.
\end{equation}

\noindent as a new feature of the data. Then, we can plot a histogram for the density of a relative uncertainty value across the 47 real observations and compare it to common probability distributions as shown by Figure~\ref{fig:uncertainty-distribution-fits}. There are two ways of extrapolating data from this histogram of real relative uncertainties in $H(z)$.

\subsubsection{Distribution Fits}\label{subsubsec:distribution-fits}

Immediately, we can see that it is not a normal distribution, and the shape is heavily skewed to the right. As a result, this study tried fitting some other common probability distributions – including log-normal, Gamma, exponential, Beta, and Weibull\_min – and evaluated the best distribution using the Kolmogorov--Smirnov (KS) test. Quantitatively, the test statistic is defined to be
\begin{equation}\label{eq:KSstat}
D_n=\sup_x\lvert F_n(x)-F(x)\rvert
\end{equation}
where $F_n(x)$ and $F(x)$ are the cumulative distribution function (CDF) of the empirical data and that of the fitted distribution respectively. From this, we can set the null hypothesis $H_0$ (not to be confused with the Hubble constant at $z=0$) for a distribution to be that the sample comes from that particular distribution. We find the following for the probability of getting a test statistic that is at least as extreme as $D_n$ assuming the null hypothesis is true:
\begin{equation}\label{eq:KS-p}
p=\mathbb{P}\big(D_n\ge \text{observed }D_n \mid H_0\big)=Q_{KS}\big(D_n\sqrt{n}\big).
\end{equation}
where
\begin{equation}\label{eq:KS-series}
Q_{KS}(x)=2\sum_{k=1}^{\infty}(-1)^{k-1}e^{-2k^2x^2}.
\end{equation}
and $n$ is the number of samples.

To prevent looping on forever, the practical algorithm will implement an early stopping condition. Let $Q_t\in[0,1]$, $Q^\ast=\max_{s\le t}Q_s$ (note $s$ is just a dummy variable to represent iteration $0\le s \le t$), $t^\ast=\arg\max_{s\le t}Q_s$, $p\in\mathbb{N}$, and $\delta\ge 0$ be the KS value at iteration $t$, the best $p$-value up to iteration $t$, the iteration where $Q^\ast$ is achieved, the patience parameter (number of allowed non-improving steps), and the optimal tolerance threshold, respectively. Then, the program stops at iteration $T=k$ in this case if and only if

\begin{equation}\label{eq:early-stop}
Q_t\le Q^\ast+\delta \;\wedge\; \forall t\in[t^\ast+1,T] \;\wedge\; T-t^\ast\ge p.
\end{equation}

Using this algorithm, the results of KS test for the mentioned distribution candidates are shown in Table~\ref{tab:ks-pvalues}.

\begin{deluxetable}{ccccc}
\tablecaption{KS-test $p$-values by distribution\label{tab:ks-pvalues}}
\tablewidth{0pt}
\tabletypesize{\footnotesize}
\tablehead{
  \colhead{\texttt{weibull\_min}} &
  \colhead{Beta} &
  \colhead{Log-normal} &
  \colhead{Exponential} &
  \colhead{Gamma}
}
\startdata
0.8404 & 0.4194 & 0.4172 & 0.0990 & 0.0001 \\
\enddata
\end{deluxetable}

Here, we see that the weibull\_min distribution has the highest $p$-value, which means it is the closest to the actual relative uncertainty distribution. Mathematically, the weibull\_min distribution is:
\begin{equation}\label{eq:weibull-pdf}
f(x)=
\begin{cases}
\frac{c}{\lambda}\,\bigl(\tfrac{x-\mu}{\lambda}\bigr)^{c-1}
\exp\!\bigl[-\bigl(\tfrac{x-\mu}{\lambda}\bigr)^{c}\bigr], & x\ge \mu,\\[4pt]
0, & x<\mu.
\end{cases}
\end{equation}
where $\lambda$, $c$, and $\mu$ are the scale parameter, shape parameter, and location parameter respectively. Therefore, we will sample relative uncertainty values from the weibull\_min distribution such that the probability of a value getting chosen is proportional to the density of the value in the distribution.

Let $x_1, x_2, \ldots, x_n$ be the $n$ real values for the relative uncertainty. To find the values of $\lambda$, $c$, and $\mu$, we will let the likelihood function be:

\begin{equation}\label{eq:likelihood}
\begin{aligned}
\mathcal{L}(c,\mu,\lambda)
&= \prod_{i=1}^{n} f(x_i;c,\mu,\lambda)\\
&= \prod_{i=1}^{n} \frac{c}{\lambda}\left(\frac{x_i-\mu}{\lambda}\right)^{c-1}
   \exp\!\left[-\left(\frac{x_i-\mu}{\lambda}\right)^{c}\right].
\end{aligned}
\end{equation}

Next, when fitting the weibull\_min distribution, we will take the log-likelihood of the above likelihood function and simplify it to get:

\begin{equation}\label{eq:loglik}
\begin{aligned}
\log \mathcal{L}(c,\mu,\lambda)
&= \sum_{i=1}^{n}\!\Big[\log c - \log \lambda + (c-1)\log\!\big(\tfrac{x_i-\mu}{\lambda}\big) \\
&\quad - \big(\tfrac{x_i-\mu}{\lambda}\big)^c\Big] = n\log c - n\log \lambda \\
&\quad + \space (c-1)\sum_{i=1}^{n}\!\log\!\big(\tfrac{x_i-\mu}{\lambda}\big) - \sum_{i=1}^{n}\!\big(\tfrac{x_i-\mu}{\lambda}\big)^c .
\end{aligned}
\end{equation}

This study then uses a Maximum Likelihood Estimation (MLE) algorithm to fit the distribution as follows:
\begin{equation}\label{eq:mle}
\begin{aligned}
(\hat{c},\hat{\mu},\hat{\lambda})
&=\arg\max_{\substack{c>0,\;\lambda>0,\\ \mu<\min x_i}}
\log\mathcal{L}(c,\mu,\lambda) \\
&=\arg\max_{\substack{c>0,\;\lambda>0,\\ \mu<\min x_i}}
\Big[
  n\log c - n\log\lambda \\
  &\quad + (c-1)\sum_{i=1}^n \log\!\Big(\tfrac{x_i-\mu}{\lambda}\Big)
  - \sum_{i=1}^n \Big(\tfrac{x_i-\mu}{\lambda}\Big)^{c}
\Big].
\end{aligned}
\end{equation}
From this, we get $c = 0.994$, $\mu = 0$, and $\lambda = 0.151$ for the relative uncertainty distribution of the 47 real data points. Given these parameters’ values, we can now randomly sample $\epsilon \sim \mathrm{Weibull}_{\min}(c,\mu,\lambda) = \mathrm{Weibull}_{\min}(0.994, 0, 0.151)$.

\subsubsection{Kernel Density Estimation (KDE)}\label{subsubsec:KDE}

Let $\epsilon_i \equiv \sigma_{H, rel}(z_i)>0$ denote the relative uncertainty attached to the $i$th real observation, and define the stabilized variable
\[
y_i \;=\; \log \epsilon_i\in\mathbb{R}.
\]
We fit a one–dimensional kernel density estimator (KDE) to $\{y_i\}_{i=1}^m$ using a Gaussian kernel $K(u)=\tfrac{1}{\sqrt{2\pi}}e^{-u^2/2}$:
\begin{equation}\label{eq:kde}
\widehat f_h(y)
=\frac{1}{m h\sqrt{2\pi}}\sum_{i=1}^m \exp\!\left(-\frac{(y-y_i)^2}{2h^2}\right),
\end{equation}
with bandwidth $h>0$ chosen by leave-one-out log-likelihood cross-validation (LCV):
\begin{equation}\label{eq:lcv}
\mathrm{LCV}(h)
=\sum_{i=1}^m \log \widehat f_{-i,h}(y_i)
\end{equation}

\noindent where

\begin{equation}\label{eq:lcv_f}
\qquad
\widehat f_{-i,h}(y)
=\frac{1}{(m-1)h\sqrt{2\pi}}\!\sum_{j\neq i}\!
\exp\!\left(-\frac{(y-y_j)^2}{2h^2}\right).
\end{equation}

From this, we found the best value to be $h=0.383961$. Note that sampling from Eq.~\eqref{eq:kde} is straightforward because it is a uniform mixture of Gaussians:
pick $I\sim\mathrm{Unif}\{1,\dots,m\}$, draw $y^\star\sim \mathcal{N}(y_I,h^2)$, and set $\epsilon^\star=e^{y^\star}$ to map back to the positive domain. To guard against pathological tails from very small samples, we clip $\epsilon^\star$ to the empirical interval $[q_{0.5\%},\,q_{99.5\%}]$ of $\{\epsilon_i\}$. Note that clipping affects only a tiny fraction of draws and stabilizes extreme tails and ablations without clipping were also checked. All KDE fitting and evaluations in this paper are performed on the $z\le 1$ sub-sample, so the learned noise law reflects the low-$z$ regime used for training the networks.

\subsubsection{Comparison of Noise Models}\label{subsubsec:noise-models-comparison}

After quantitatively comparing the parametric $\texttt{weibull}_\texttt{min}$ fit (the best fit found out of all the distributions examined based on $p$-value) on $\epsilon$ against the nonparametric KDE on $y=\log \epsilon$ through another KS test (Eq~\eqref{eq:KSstat}) between the real $H(z)$ relative uncertainty distribution and the proposed noise models, this study chose to sample $\epsilon$ using KDE (Section~\ref{subsubsec:KDE}) instead of $\texttt{weibull}_\texttt{min}$ (Section~\ref{subsubsec:distribution-fits}) since KDE has a higher $p$-value, as shown in Table~\ref{tab:ks-pvals-noise-models}.

\begin{deluxetable}{cc}
\tablecaption{KS $p$-values for KDE and $\texttt{weibull}_\texttt{min}$ noise models\label{tab:ks-pvals-noise-models}}
\tablewidth{0pt}
\tabletypesize{\footnotesize}
\tablehead{
  \colhead{KDE} & \colhead{\texttt{weibull\_min}}
}
\startdata
0.902129 & 0.389292 \\
\enddata
\end{deluxetable}

\subsection{Synthetic Data Generation}\label{subsec:synthetic-data-generation}

To generate clean synthetic Hubble expansion graphs, this paper first sampled 100,000 sets of values for cosmological parameters through the conditions shown in Table~\ref{tab:priors}. Then, we can input the selected cosmological parameters into the first Friedmann equation (Eq.~\eqref{eq:friedmann-simplified}) at randomly chosen points of $z$ to generate $H_{\text{clean}}(z)$.

\begin{deluxetable}{lll}
\tablecaption{Model parameters and priors for flat $\Lambda$CDM ($z\le1$).\label{tab:priors}}
\tablewidth{0pt}
\tabletypesize{\footnotesize}
\tablehead{
  \colhead{Parameter} & \colhead{Distribution} & \colhead{Prior}
}
\startdata
$\Omega_{m,0}$        & Uniform        & $[0,1]$ \\
$\Omega_{r,0}$        & Fixed          & $0$ \\
$\Omega_{\Lambda,0}$  & Deterministic  & $1-\Omega_{m,0}$ \\
$\Omega_{k,0}$        & Fixed          & $0$ \\
$w$                   & Fixed          & $-1$ \\
\enddata
\tablecomments{Assumes flat $\Lambda$CDM with negligible radiation at low $z$.}
\end{deluxetable}

Now, from Section~\ref{subsubsec:noise-models-comparison}, we will sample each $\epsilon_i$ from the Gaussian KDE in Section~\ref{subsubsec:KDE} as our relative uncertainty noise for data point \emph{i}. Then, we can define $\Delta H_i$ to be $\epsilon_iH(z_i)$ for data point $i$. From here, we can sample
\begin{equation}\label{eq:gaussian-noise-eta}
\eta_i \sim \mathcal{N}\!\big(0, k^2\Delta H_i^{\,2}\big),
\end{equation}
and then add noise vertically using the sampled $\eta_i$ using Eq.~\eqref{eq:gaussian-noise-generation}
\begin{equation}\label{eq:gaussian-noise-generation}
H_{\text{noisy}}(z_i) = H_{\text{clean}}(z_i) + \eta_i.
\end{equation}

Here, $k$ is a coefficient in front of $\Delta H_i^2$ that determines the SNR of the synthetic data. Specifically, we see that higher $k$ increases the standard deviation for $\eta_i$, leading to a lower SNR. In other words, it can be shown that $\text{SNR} \propto \frac{1}{k}$. In this paper, we generated 3 different synthetic training data, one with $k=0.5$ (high SNR; two times higher than normal SNR), one with $k=1$ (normal SNR), and one with $k=2$ (low SNR; two times smaller than normal SNR). Note that theoretically, for the synthetic data generated to match the patterns in the real data as much as possible, we should use $k=1$; however, this paper tested the other values for $k$ as an experiment to evaluate how different SNR training data would affect the performance of each ANN model architecture, as stated in Section~\ref{sec:introduction}.

\subsection{ANN Models}\label{subsec:ann-models}

\subsubsection{Data Preprocessing}\label{sec:data-preprocessing}
In total, this study used the synthetic data generation algorithm outlined in Section~\ref{subsec:synthetic-data-generation} to generate 100,000 synthetic Hubble expansion graphs, each with 100 observations to use as the training and validation data for the ANN models. 

However, to generate more training data and match the input shape (47 Hubble expansion observations) of the real data provided by \citet{Bouali2023}, we applied bootstrap sampling to obtain accurate predictions from ANNs on the real data (which is done in Section~\ref{subsec:architecture-predictions-and-evaluation}). Specifically, from the original sample of 100 observations, we generated 10 bootstrap replicates $X^{\ast (1)},  X^{\ast (2)}, \ldots , X^{\ast (10)}$, each consisting of 47 observations drawn with replacement from the original 100-observation sample. Formally, the process can be written as:
\begin{equation}
\begin{aligned}\label{eq:bootstrap}
&\bigl\{X^{\ast (1)}, \ldots , X^{\ast (10)}\bigr\}
= \Bigl\{
\{\,x^{(b)}_{i_1}, \ldots , x^{(b)}_{i_{47}}\,\}_{j=1}^{47}
\;\Big|\; \\
&\quad i_j \sim \text{iid } \mathrm{Uniform}(\{1,\ldots,100\})
\Bigr\}_{b=1}^{10}.
\end{aligned}
\end{equation}

After bootstrapping, we now have 1,000,000 samples, which the models will then use for training and validation. In particular, we will divide the data so that they contain 80\% training data, 10\% validation data, and 10\% test data.  

\subsubsection{Model Architectures}\label{subsubsec:model-architectures}

As stated in Section~\ref{sec:introduction}, this paper tests 5 different ANN architectures that are more complex than ParamANN. In particular, we will be examining the performance of multiplayer perceptrons (MLP), convolutional neural networks (CNN), Bidirectional Long Short-Term Memory (BiLSTM) recurrent neural networks (RNN), CNN + BiLSTM RNN, and CNN + Gated Recurrent Unit (GRU) RNN models, with their specific architectures shown in Table~\ref{tab:all-model-architectures}.

\begin{deluxetable*}{cllrr}
\tablecaption{Neural network architectures: layer configurations and parameter counts for all models.\label{tab:all-model-architectures}}
\tablewidth{0pt}
\tabletypesize{\scriptsize}
\tablehead{
\colhead{\#} & \colhead{Layer type} & \colhead{Hyperparameters} & \colhead{Output units} & \colhead{Params}
}
\startdata
\cutinhead{MLP}
0 & Input            & shape $(47,2)$   & $(47,2)$ & 0 \\
1 & Flatten          & ---              & $94$     & 0 \\
2 & Dense            & units=64, ReLU   & $64$     & 6,080 \\
3 & Dense            & units=32, ReLU   & $32$     & 2,080 \\
4 & Dense (output)   & units=2          & $2$      & 66 \\
\tableline
\multicolumn{4}{r}{Total trainable parameters} & 8,226 \\
\cutinhead{CNN}
0 & Input              & shape $(47,2)$                            & $(47,2)$  & 0 \\
1 & Conv1D             & filters=64, kernel=3, padding=same, ReLU  & $(47,64)$ & 448 \\
2 & Conv1D             & filters=64, kernel=3, padding=same, ReLU  & $(47,64)$ & 12,352 \\
3 & GlobalMaxPooling1D & ---                                        & $64$      & 0 \\
4 & Dense              & units=128, ReLU                            & $128$     & 8,320 \\
5 & Dense              & units=64, ReLU                             & $64$      & 8,256 \\
6 & Dense (output)     & units=2                                    & $2$       & 130 \\
\tableline
\multicolumn{4}{r}{Total trainable parameters} & 29,506 \\
\cutinhead{RNN (BiLSTM)}
0 & Input              & shape $(47,2)$                        & $(47,2)$   & 0 \\
1 & Bidirectional LSTM & units=96, return\_sequences=True      & $(47,192)$ & 76,032 \\
2 & Bidirectional LSTM & units=64                              & $128$      & 131,584 \\
3 & Dense              & units=128, ReLU                       & $128$      & 16,512 \\
4 & Dense (output)     & units=2                               & $2$        & 258 \\
\tableline
\multicolumn{4}{r}{Total trainable parameters} & 224,386 \\
\cutinhead{CNN + RNN (BiLSTM)}
0 & Input                  & shape $(47,2)$                                 & $(47,2)$   & 0 \\
1 & Conv1D                 & filters=64, kernel=5, padding=same, ReLU       & $(47,64)$  & 704 \\
2 & Conv1D                 & filters=64, kernel=5, padding=same, ReLU       & $(47,64)$  & 20,544 \\
3 & Bidirectional LSTM     & units=64, return\_sequences=True               & $(47,128)$ & 66,048 \\
4 & GlobalAveragePooling1D & ---                                             & $128$      & 0 \\
5 & Dense                  & units=128, ReLU                                 & $128$      & 16,512 \\
6 & Dense                  & units=64, ReLU                                  & $64$       & 8,256 \\
7 & Dense                  & units=32, ELU                                   & $32$       & 2,080 \\
8 & Dense (output)         & units=2                                         & $2$        & 66 \\
\tableline
\multicolumn{4}{r}{Total trainable parameters} & 114,210 \\
\cutinhead{CNN + RNN (GRU)}
0 & Input              & shape $(47,2)$                            & $(47,2)$  & 0 \\
1 & Conv1D             & filters=64, kernel=3, padding=same, ReLU & $(47,64)$ & 448 \\
2 & Conv1D             & filters=64, kernel=3, padding=same, ReLU & $(47,64)$ & 12,352 \\
3 & GRU                & units=128                                 & $128$     & 74,496 \\
4 & Dense              & units=128, ReLU                           & $128$     & 16,512 \\
5 & Dense (output)     & units=2                                   & $2$       & 258 \\
\tableline
\multicolumn{4}{r}{Total trainable parameters} & 104,066 \\
\enddata
\end{deluxetable*}

\subsubsection{Model Training and Output}\label{subsubsec:model-training-output}

The CNN and MLP model stated in Section~\ref{subsubsec:model-architectures} will be trained 100 different times, and the other models will be trained 20 times with different random initial states over 100 epochs with the mean squared error (MSE) as its loss function, for each synthetic training dataset (low SNR, medium SNR, high SNR) generated in Section~\ref{subsec:synthetic-data-generation}. Formally, MSE is defined as

\begin{equation}\label{eq:mse}
\mathrm{MSE}=\frac{1}{N}\sum_{i=1}^{N}\bigl(y_i-\hat{y}\bigr)^2.
\end{equation}

For more accurate and flexible backpropagation, this model uses the Adam optimizer from the Keras Python library, which combines the advantages of Momentum and RMSProp. In addition, to speed up the training process, we have also deployed an early-stopping algorithm (identical to that of Eq.~\eqref{eq:early-stop}) to stop the model training if it is no longer statistically improving the model predictions. 

After training 100 instances (with different random initial states) of the MLP and CNN model and 20 instances for the other model architectures (as stated in Section~\ref{subsubsec:model-architectures}), this paper will find the mean and standard deviation of its outputs using the standard statistics formulas:

 \begin{equation}\label{eq:mean}
\mu_{\Omega_p} = \frac{1}{N}\sum_{i=1}^{N}{\Omega_{p, i}}
\end{equation}

\noindent and

\begin{equation}\label{eq:std}
\sigma_{\Omega_p} = \sqrt{\frac{\sum_{i=1}^{N}{(\Omega_{p, i} - \mu_{\Omega_p})^2}}{N-1}},
\end{equation}

\noindent for each density parameter $p$, where $p=m, \Lambda$ in this case.

\subsection{Evaluation Metrics}\label{subsec:evaluation-metrics}

We assessed agreement between this paper's ANN models and the real Hubble expansion data compiled by \citet{Bouali2023} with (i) a two-sample KS test on sorted \(H(z)\) values (where we used the same quantitative calculations as the KS test performed in Section~\ref{subsubsec:distribution-fits}); (ii) the 1-Wasserstein distance \(W_1\) between the two \(H(z)\) distributions (described in Section~\ref{subsubsec:wasserstein}); (iii) a $\chi^2$ statistic using observational uncertainties only for \(\nu=N\) and a more conservative \(\nu=N-2\) (described in Section~\ref{subsubsec:chi2-obs}); and (iv) a generalized least squares (GLS) chi-square that accounts for parameter-induced and correlated uncertainty for \(\nu=N\) and a more conservative \(\nu=N-2\) (described in Section~\ref{subsubsec:chi2-gls}). Note that for KS and $W_1$, we evaluated $H_{\rm model}(z)$ on the observed redshift grid $\{z_i\}_{i=1}^N$ so that both samples have equal size $N=47$.

Afterwards, we used the ranking system stated in Section~\ref{subsubsec:ranking} to evaluate each model, determining which architecture and SNR performed the best and worst based on these metrics. 

\subsubsection{1-Wasserstein Distance}\label{subsubsec:wasserstein}
The 1-Wasserstein Distance quantifies the overall discrepancy between the empirical distributions of the model and observed $H(z)$ values, where smaller values are better. This metric complements KS by aggregating differences across the entire distribution rather than focusing on a single worst-case deviation.

Formally, for equal sample sizes $N$, let $H_{\rm model}^{\uparrow}(i)$ and $H_{\rm obs}^{\uparrow}(i)$ denote the $i$th order statistics (sorted ascending). Then, the empirical 1-Wasserstein distance is

\begin{equation}\label{eq:wasserstein-distance}
\begin{aligned}
W_1 &= \frac{1}{N}\sum_{i=1}^{N} \big| H_{\rm model}^{\uparrow}(i) - H_{\rm obs}^{\uparrow}(i) \big| \\
    &= \int_{\mathbb{R}} \big|F_{\rm model}(x)-F_{\rm obs}(x)\big|\,dx,
\end{aligned}
\end{equation}

\noindent where $F_{\rm model}$ and $F_{\rm obs}$ are the empirical CDFs. Therefore, note that in this case $W_1$ has the same units as $H$ (km\,s$^{-1}$\,Mpc$^{-1}$). Additionally, it is sensitive to location and scale differences, thus accumulating discrepancies across quantiles.

\subsubsection{Observation Only $\chi^2$ Statistic}\label{subsubsec:chi2-obs}
A $\chi^2$ test using only the uncertainty in observational $\sigma_H$ from \citet{Bouali2023} tests pointwise agreement between $H_{\rm model}(z)$ and $H_{\rm obs}(z)$, serving as a statistical goodness-of-fit test.

Quantitatively, with the observational standard deviations $\sigma_H(z_i)$ and residuals $r_i=H_{\rm obs}(z_i)-H_{\rm model}(z_i)$,
\begin{equation}
\chi^2_{\rm obs} \;=\; \sum_{i=1}^{N} \frac{r_i^2}{\sigma_H^2(z_i)}\,,
\qquad
\tilde{\chi}^2_{\rm obs} \equiv \frac{\chi^2_{\rm obs}}{\nu}\,,
\end{equation}
where we reported $p$-values for $\nu=N$ (parameters not fit to these data) and, conservatively, also for $\nu=N-2$.

For this metric, $\tilde{\chi}^2_{\rm obs}\!\approx\!1$ indicates a typical fit, $\gg 1$ suggests tension or underestimated errors, and $\ll 1$ suggests overestimated errors or unmodeled correlations. This formulation assumes independent observational errors (diagonal covariance) and does not include model-parameter uncertainty.

From here, the $\chi^2$ test $p$-value is simply the survival function of the $\chi^2$ distribution,

\begin{equation}\label{eq:chi-squared-p-value}
p \;=\; \Pr\!\big[\chi^2_\nu \ge \chi^2_{\text{obs}}\big]
= 1 - F_{\chi^2_\nu}\!\big(\chi^2_{\text{obs}}\big)
= Q\!\left(\frac{\nu}{2},\, \frac{\chi^2_{\text{obs}}}{2}\right),
\end{equation}

\noindent where $Q(a, x)$ is the regularized incomplete gamma function. The $p$-value in this case represents the probability, assuming the model is correct and that the stated error model holds, that random noise would produce a $\chi^2$ at least as large as the one observed.

\subsubsection{GLS $\chi^2$ Statistic}\label{subsubsec:chi2-gls}

Unlike the observation only $\chi^2$ statistic described in Section~\ref{subsubsec:chi2-obs}, a GLS $\chi^2$ statistic incorporates parameter-induced and correlated uncertainty in $H_{\rm model}(z)$ by marginalizing linear shifts due to $(\Omega_m,\Omega_\Lambda)$. Therefore, this metric typically yields a less stringent (larger) $p$ than obs-only $\chi^2$ because plausible parameter variations are accounted for.

Let $\mathbf{r}\in\mathbb{R}^N$ collect residuals, $\mathbf{C}_{\rm obs}=\mathrm{diag}(\sigma_H^2(z_i))$, and

\begin{equation}
\mathbf{C}_{\rm model}(z_i,z_j) \;=\; \mathbf{J}(z_i)\,\boldsymbol\Sigma_{\Omega}\,\mathbf{J}(z_j)^\top,
\end{equation}

\begin{equation}
\mathbf{J}(z)=\begin{bmatrix}\partial H/\partial\Omega_m & \partial H/\partial\Omega_\Lambda\end{bmatrix}_{(\bar\Omega_m,\bar\Omega_\Lambda)}.
\end{equation}

\noindent Then, formally, the generalized least-squares statistic is defined as
\begin{equation}
\chi^2_{\rm GLS} \;=\; \mathbf{r}^\top \big(\mathbf{C}_{\rm obs}+\mathbf{C}_{\rm model}\big)^{-1}\mathbf{r}\,,
\end{equation}
with $p$-values reported for $\nu=N$ and (conservatively) $\nu=N-2$. An equivalent Schur-Woodbury form,

\begin{equation}
\begin{split}
\chi^2_{\rm GLS} \;=&\; \mathbf{r}^\top \mathbf{C}_{\rm obs}^{-1}\mathbf{r}\;-\; \\
& \big(\mathbf{J}^\top \mathbf{C}_{\rm obs}^{-1}\mathbf{r}\big)^\top
\Big(\boldsymbol\Sigma_{\Omega}^{-1}+\mathbf{J}^\top \mathbf{C}_{\rm obs}^{-1}\mathbf{J}\Big)^{-1}
\big(\mathbf{J}^\top \mathbf{C}_{\rm obs}^{-1}\mathbf{r}\big),
\end{split}
\end{equation}

\noindent avoids inverting an $N\times N$ matrix.

From this definition, we can see $\boldsymbol\Sigma_{\Omega}$ is estimated as the sample covariance of $(\Omega_m,\Omega_\Lambda)$ across independent runs. Note that the GLS construction relies on a local linearization of $H(z;\Omega_m,\Omega_\Lambda)$ and a Gaussian approximation for parameter uncertainty; however, if a posterior covariance is available, it should be used in place of the sample covariance. By construction, it can be shown that $\chi^2_{\rm GLS}\!\le\!\chi^2_{\rm obs}$.

The $p$-value for this test is defined identically to the definition in Section~\ref{subsubsec:chi2-obs}, since both metrics are $\chi^2$ goodness-of-fit tests.

\subsubsection{Model Ranking}\label{subsubsec:ranking}

To avoid over-weighting correlated $\chi^2$ metrics, we aggregate metrics by family.
We compute a distributional rank $R_{\rm dist}=\tfrac{1}{2}\{\mathrm{rank}(\text{KS }p) + \mathrm{rank}(W_1\ \text{ascending})\}$ and a pointwise rank $R_{\chi^2}=\tfrac{1}{2}\{\mathrm{rank}(p_{\chi^2}^{\mathrm{obs}}(\nu=N)) + \mathrm{rank}(p_{\chi^2}^{\mathrm{GLS}}(\nu=N))\}$ using competition ranking (ties share a rank). The final score is the average $\bar R=\tfrac{1}{2}(R_{\rm dist}+R_{\chi^2})$, with smaller $\bar R$ indicating better overall performance. Note that the $p$-value metric ranks are in descending order because a higher $p$-value indicates a stronger model fit to real data compiled by \citet{Bouali2023}, while $W_1$ is ranked in ascending order because a lower Wasserstein Distance means less distributional difference between the model and the real data, which thus indicates a more accurate model.

\section{Results and Discussion}\label{sec:results-and-discussion}

\subsection{Architecture Predictions and Evaluation}\label{subsec:architecture-predictions-and-evaluation}

The results and their respective evaluation metrics stated in Section~\ref{subsec:evaluation-metrics} for each model architecture and SNR are shown in Table~\ref{tab:model-results}. In general, we found that for KS tests and $\chi^2$ goodness-of-fit tests, most models have $p\text{-value}>0.05$, representing that the model is consistent with the real data compiled by \citet{Bouali2023} given the error model. In addition, the results show that different SNRs did not influence the model predictions' standard deviation significantly. From these evaluation metrics, we used the ranking procedure described in Section~\ref{subsubsec:ranking} to make an accuracy (in terms of its fit with real data compiled by \citet{Bouali2023}) leaderboard displaying the rankings for each model architecture and SNR, as shown in Table~\ref{tab:models-leaderboard}.

\begin{deluxetable*}{ccccccccccccc}\label{tab:model-results}
\tablecaption{Results for all models: $\Omega_m$, $\Omega_\Lambda$, Test MSE, KS $p$, raw $\chi^2$ (obs) and $\chi^2$ (GLS), with $p$-values for $\nu{=}N$ and $\nu{=}N{-}2$, and Wasserstein for $0<z\le1$.}
\tablehead{\colhead{$k$} & \colhead{$\Omega_m$} & \colhead{$\Omega_{\Lambda}$} & \colhead{Test MSE} & \colhead{KS $p$} & \colhead{$\chi^2$ (obs)} & \colhead{$\chi^2$ $p_N$ (obs)} & \colhead{$\chi^2$ $p_{N-2}$ (obs)} & \colhead{$\chi^2$ (GLS)} & \colhead{$\chi^2$ $p_N$ (GLS)} & \colhead{$\chi^2$ $p_{N-2}$ (GLS)} & \colhead{Wasserstein}}
\startdata
\cutinhead{MLP}
0.5 & 0.255025 $\pm$ 0.016371 & 0.744008 $\pm$ 0.017245 & 0.001460 $\pm$ 0.000285 & 0.466 & 62.00 & 0.070 & 0.047 & 39.24 & 0.783 & 0.714 & 2.911603 \\
1.0 & 0.260882 $\pm$ 0.031540 & 0.738979 $\pm$ 0.032244 & 0.003866 $\pm$ 0.008070 & 0.466 & 55.28 & 0.190 & 0.140 & 34.06 & 0.921 & 0.883 & 2.654717 \\
2.0 & 0.247537 $\pm$ 0.062886 & 0.752841 $\pm$ 0.063119 & 0.010515 $\pm$ 0.017037 & 0.321 & 70.53 & 0.015 & 0.009 & 33.06 & 0.938 & 0.907 & 3.200898 \\
\cutinhead{CNN}
0.5 & 0.262530 $\pm$ 0.015860 & 0.737210 $\pm$ 0.016189 & 0.000353 $\pm$ 0.000107 & 0.466 & 53.70 & 0.233 & 0.175 & 37.95 & 0.824 & 0.763 & 2.594975 \\
1.0 & 0.257648 $\pm$ 0.013035 & 0.742436 $\pm$ 0.012873 & 0.000858 $\pm$ 0.000250 & 0.466 & 58.57 & 0.120 & 0.084 & 41.22 & 0.710 & 0.633 & 2.781500 \\
2.0 & 0.240736 $\pm$ 0.009474 & 0.759299 $\pm$ 0.009234 & 0.002037 $\pm$ 0.000134 & 0.210 & 80.24 & 0.002 & 0.001 & 55.41 & 0.187 & 0.137 & 3.519007 \\
\cutinhead{RNN (BiLSTM)}
0.5 & 0.276381 $\pm$ 0.005404 & 0.723194 $\pm$ 0.004783 & 0.000143 $\pm$ 0.000024 & 0.812 & 42.43 & 0.662 & 0.582 & 39.70 & 0.766 & 0.696 & 2.098459 \\
1.0 & 0.267900 $\pm$ 0.004297 & 0.731387 $\pm$ 0.004161 & 0.000436 $\pm$ 0.000026 & 0.466 & 48.97 & 0.394 & 0.317 & 45.91 & 0.518 & 0.434 & 2.401745 \\
2.0 & 0.261558 $\pm$ 0.003751 & 0.738595 $\pm$ 0.003231 & 0.001511 $\pm$ 0.000015 & 0.466 & 54.48 & 0.211 & 0.157 & 50.80 & 0.326 & 0.255 & 2.624033 \\
\cutinhead{CNN+RNN (BiLSTM)}
0.5 & 0.270499 $\pm$ 0.005467 & 0.729688 $\pm$ 0.005613 & 0.000165 $\pm$ 0.000027 & 0.466 & 46.51 & 0.493 & 0.410 & 43.04 & 0.638 & 0.556 & 2.290337 \\
1.0 & 0.264687 $\pm$ 0.006041 & 0.735158 $\pm$ 0.006076 & 0.000512 $\pm$ 0.000108 & 0.466 & 51.61 & 0.298 & 0.231 & 46.03 & 0.513 & 0.429 & 2.512107 \\
2.0 & 0.255578 $\pm$ 0.006155 & 0.744547 $\pm$ 0.006149 & 0.001587 $\pm$ 0.000060 & 0.466 & 60.84 & 0.085 & 0.058 & 52.12 & 0.282 & 0.217 & 2.867496 \\
\cutinhead{CNN+RNN (GRU)}
0.5 & 0.272747 $\pm$ 0.005457 & 0.727459 $\pm$ 0.005846 & 0.000152 $\pm$ 0.000027 & 0.639 & 44.79 & 0.565 & 0.481 & 41.69 & 0.692 & 0.613 & 2.209074 \\
1.0 & 0.266668 $\pm$ 0.005310 & 0.732569 $\pm$ 0.005476 & 0.000493 $\pm$ 0.000069 & 0.466 & 50.06 & 0.353 & 0.279 & 45.76 & 0.524 & 0.440 & 2.448760 \\
2.0 & 0.261927 $\pm$ 0.007230 & 0.738869 $\pm$ 0.006642 & 0.001552 $\pm$ 0.000046 & 0.466 & 53.85 & 0.229 & 0.172 & 44.64 & 0.571 & 0.487 & 2.598593 \\
\enddata
\tablecomments{KS $p$ from two-sample KS on sorted $H(z)$ samples (right-continuous EDFs); note that KS statistics are discretized due to finite $N$ (values of $m/N$). Wasserstein is the empirical $W_1$ distance between model and real $H$ values, reported in the same units as $H$ (km s$^{-1}$ Mpc$^{-1}$). $\chi^2$ (obs) uses \emph{only} observational $\sigma_H$ with $\nu{=}N$ (and we also report $\nu{=}N{-}2$ $p$-values). $\chi^2$ (GLS) uses the full covariance $C_{\rm tot}=\mathrm{diag}(\sigma_H^2)+J\,\Sigma_\Omega\,J^\top$, where $J=[\partial H/\partial\Omega_m,\partial H/\partial\Omega_\Lambda]$ at the mean parameters and $\Sigma_\Omega$ is the sample covariance across runs; $p$-values shown for both $\nu{=}N$ and $\nu{=}N{-}2$.}
\end{deluxetable*}

From Table~\ref{tab:models-leaderboard}, we see that higher SNR training data do indeed make the model perform more optimally, shown by the fact that 1) the top 4 models all have $k=0.5$; 2) the bottom 4 models all have $k=2.0$; and 3) for every model architecture, the highest $k$ performed the worst and the lowest $k$ performed the best, except for MLP. 

In addition, from the model rankings, we can conclude that the best model architecture out of the 5 tested for high SNR ($k=0.5$) and normal SNR ($k=1.0$) is RNN (BiLSTM), while the best model architecture for low SNR ($k=2.0$) is CNN + RNN (GRU). On the other hand, we can also conclude that the worst model architecture for high SNR is MLP, while the worst model architecture for normal and low SNR is CNN.

\begin{deluxetable*}{cccccc}\label{tab:models-leaderboard}
\tablecaption{Leaderboard by family-balanced average rank (Option A). $R_{\rm dist}$ averages ranks of KS $p$ (higher better) and $W_1$ (lower better). $R_{\chi^2}$ averages ranks of $\chi^2$ $p$ (obs-only, $\nu{=}N$) and $\chi^2$ $p$ (GLS, $\nu{=}N$). Smaller values are better.}
\tablehead{\colhead{$\quad$Rank$\quad$} & \colhead{Model} & \colhead{$k$} & \colhead{$R_{\rm dist}$} & \colhead{$R_{\chi^2}$} & \colhead{Avg. Rank}}
\startdata
1 & RNN (BiLSTM) & 0.5 & 1.00 & 3.00 & 2.00 \\
2 & CNN+RNN (GRU) & 0.5 & 2.00 & 4.50 & 3.25 \\
3 & CNN+RNN (BiLSTM) & 0.5 & 3.00 & 5.50 & 4.25 \\
4 & CNN & 0.5 & 5.00 & 5.00 & 5.00 \\
5 & RNN (BiLSTM) & 1.0 & 3.50 & 7.50 & 5.50 \\
6 & CNN+RNN (GRU) & 1.0 & 4.00 & 7.50 & 5.75 \\
7 & MLP & 1.0 & 6.50 & 6.00 & 6.25 \\
8 & CNN+RNN (BiLSTM) & 1.0 & 4.50 & 9.00 & 6.75 \\
9 & CNN+RNN (GRU) & 2.0 & 5.50 & 8.50 & 7.00 \\
10 & CNN & 1.0 & 7.00 & 8.50 & 7.75 \\
11 & MLP & 0.5 & 8.00 & 8.50 & 8.25 \\
12 & RNN (BiLSTM) & 2.0 & 6.00 & 11.00 & 8.50 \\
13 & CNN+RNN (BiLSTM) & 2.0 & 7.50 & 13.00 & 10.25 \\
14 & MLP & 2.0 & 14.00 & 7.50 & 10.75 \\
15 & CNN & 2.0 & 15.00 & 15.00 & 15.00 \\
\enddata
\tablecomments{Ranks use competition ranking (``min''): ties share a rank; subsequent ranks skip. Families are equally weighted to avoid overemphasizing correlated $\chi^2$ metrics as stated specifically in Section~\ref{subsubsec:ranking}.}
\end{deluxetable*}

\subsection{Comparison of Results with Past Research}

Figures~\ref{fig:posteriors} and \ref{fig:posteriors-cont} show how the normal posteriors of each SNR ($k=0.5$, $k=1.0$, and $k=2.0$) for each model architecture compares with the results of \citet{Planck2020}, which used traditional MCMC methods such as the Metropolis-Hastings algorithm, and ParamANN \citep{Pal2024}, which used a single hidden dense layer as its ANN architecture. 

We see from Table~\ref{tab:z-score-ParamANN} that all of the ANN architectures for all SNR tested in this paper are statistically consistent within 1 standard deviation ($68\%$ confidence level) of ParamANN. This demonstrates that, regardless of the ANN architecture used, the posterior distribution should not statistically deviate by significant margins. 

On the other hand, most models and SNRs are not statistically consistent with that of Planck collaboration at all, as shown in Table~\ref{tab:z-score-planck} by the fact that many posterior distribution differs from those of Planck Collaboration by more than 3 standard deviations ($>99\%$ confidence level). This does not immediately demonstrate that the results of ANN models are necessarily wrong, as discussed in Section~\ref{subsec:architecture-predictions-and-evaluation}, but rather that the novel method of ANN's results differ significantly from traditional MCMC methods' results. It is worth noting, however, that the MLP model for $k=2.0$ is within 1 standard deviation of Planck Collaboration, showing that it is still possible to create an ANN model architecture that is statistically significant within 1 standard deviation with the posterior results of traditional MCMC methods.


\begin{figure*}
\centering
\gridline{
  \fig{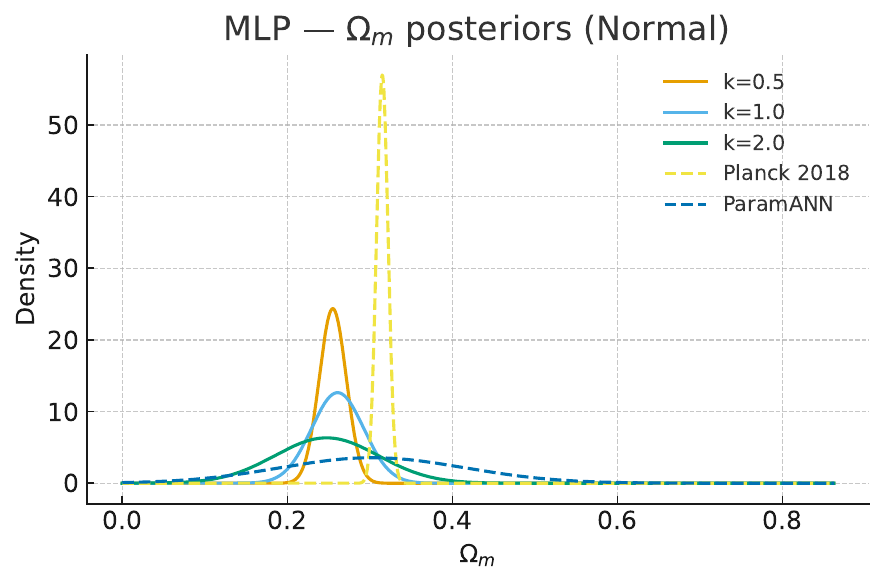}{0.47\textwidth}{(a) MLP: $\Omega_m$}
  \fig{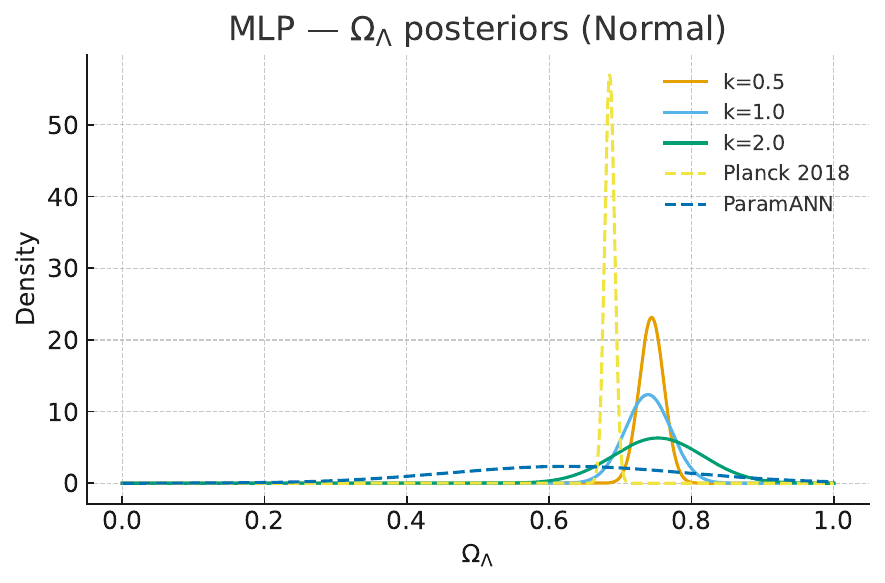}{0.47\textwidth}{(b) MLP: $\Omega_\Lambda$}
}
\gridline{
  \fig{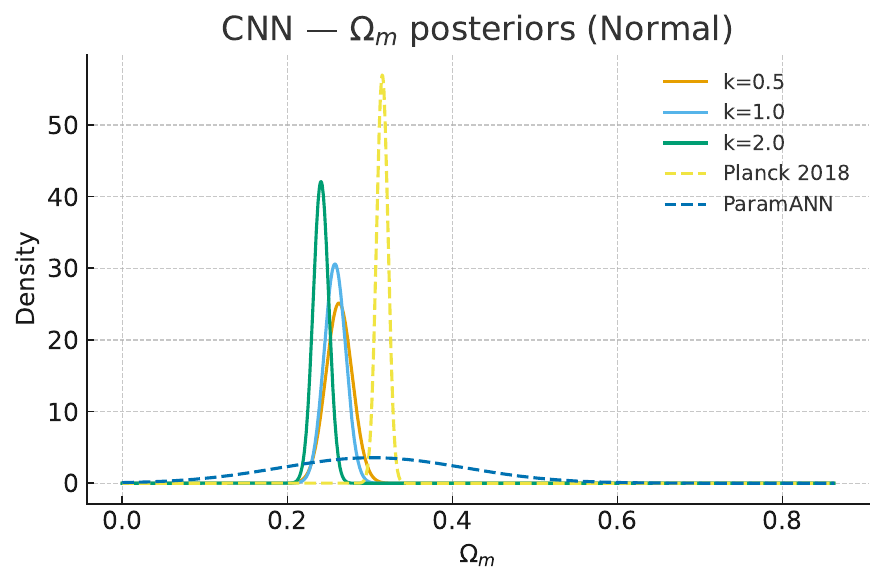}{0.47\textwidth}{(c) CNN: $\Omega_m$}
  \fig{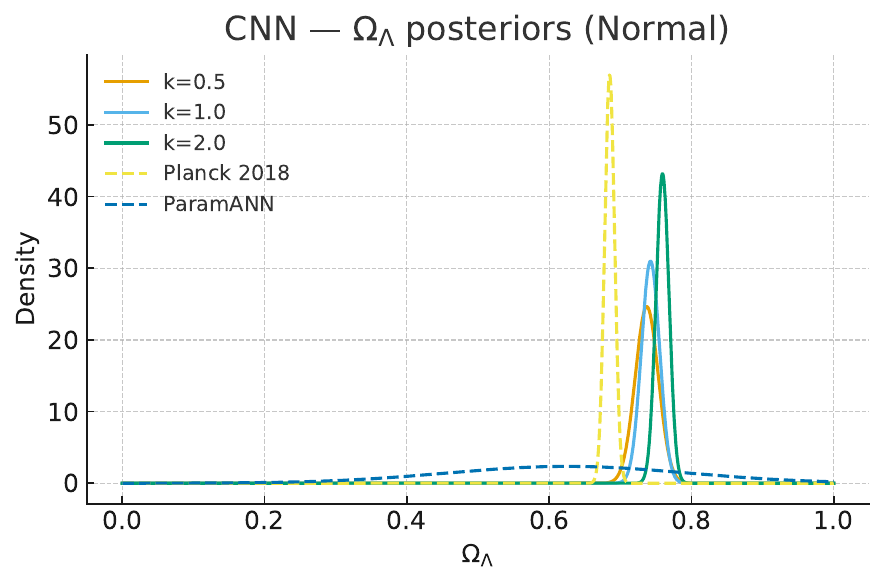}{0.47\textwidth}{(d) CNN: $\Omega_\Lambda$}
}
\gridline{
  \fig{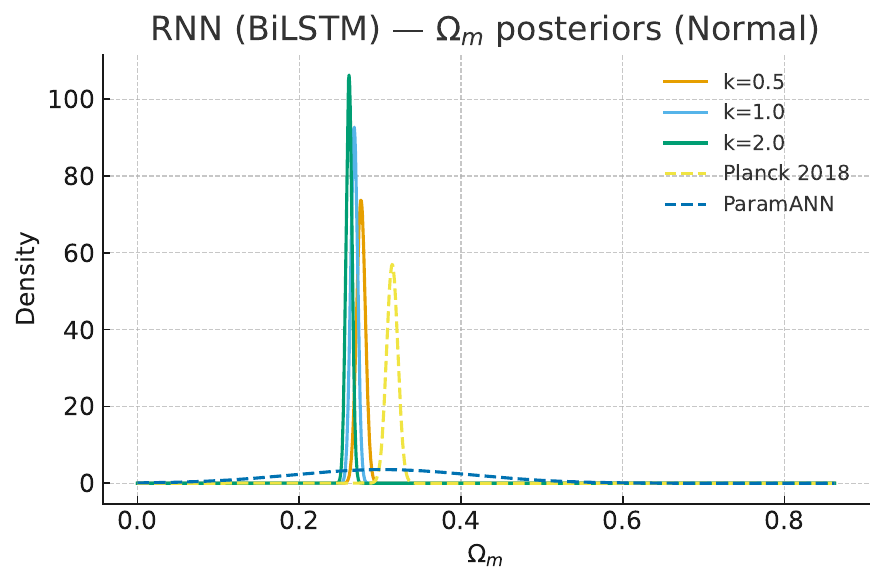}{0.47\textwidth}{(e) RNN (BiLSTM): $\Omega_m$}
  \fig{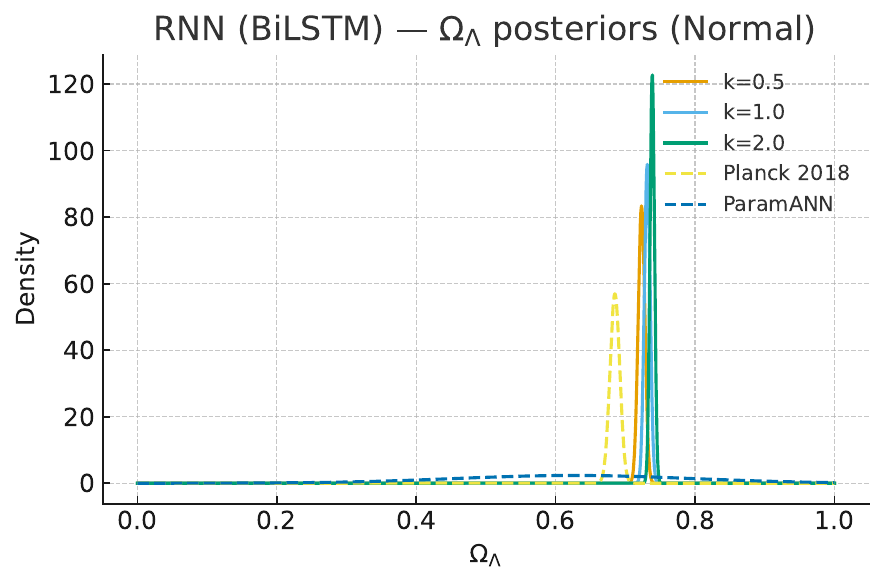}{0.47\textwidth}{(f) RNN (BiLSTM): $\Omega_\Lambda$}
}
\caption{Normalized Gaussian posteriors for each model. \textit{Solid lines:} model posteriors at $k=0.5,1.0,2.0$ (from Table~\ref{tab:model-results}). \textit{Dashed lines:} Planck 2018 and ParamANN posteriors.}
\label{fig:posteriors}
\end{figure*}

\begin{figure*}
\centering
\gridline{
  \fig{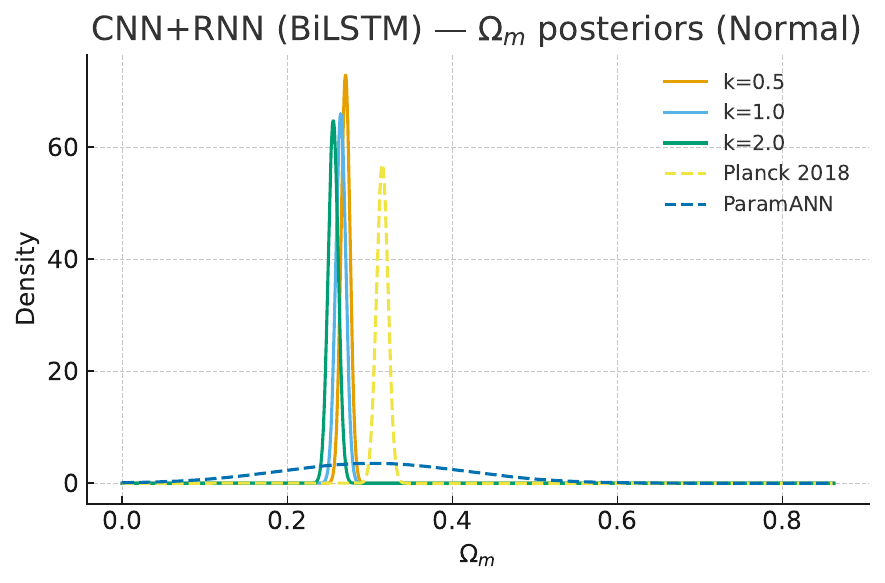}{0.47\textwidth}{(g) CNN+RNN (BiLSTM): $\Omega_m$}
  \fig{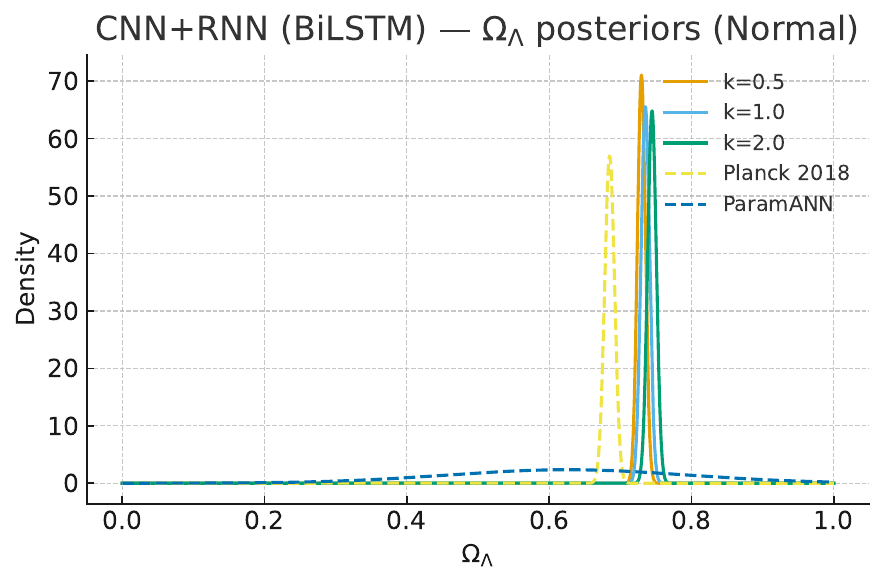}{0.47\textwidth}{(h) CNN+RNN (BiLSTM): $\Omega_\Lambda$}
}
\gridline{
  \fig{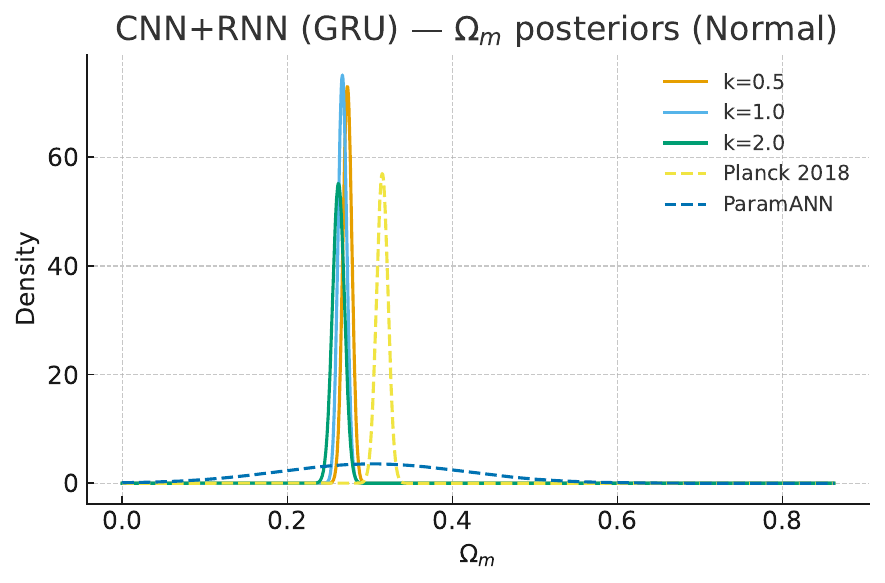}{0.47\textwidth}{(i) CNN+RNN (GRU): $\Omega_m$}
  \fig{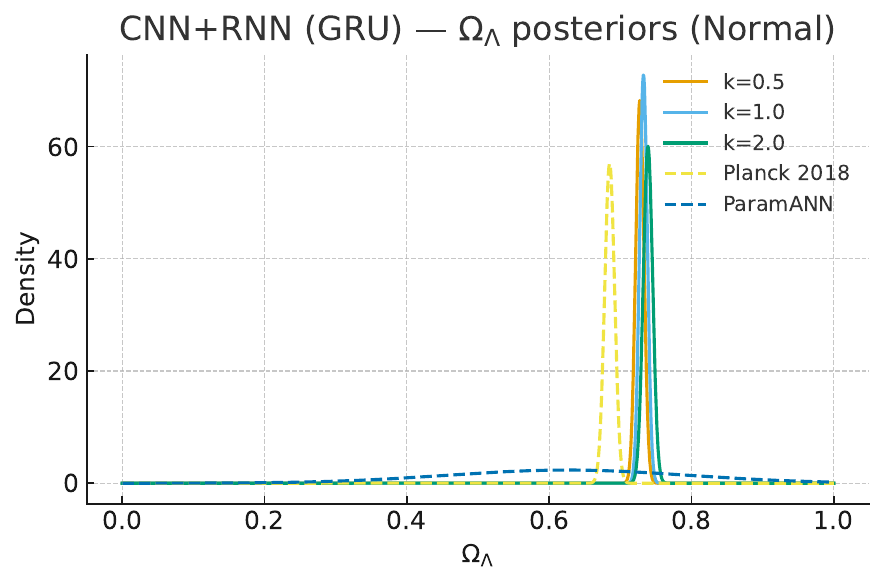}{0.47\textwidth}{(j) CNN+RNN (GRU): $\Omega_\Lambda$}
}
\caption{\textit{(Continued)} Same as Fig.~\ref{fig:posteriors}, panels (g)–(j).}
\label{fig:posteriors-cont}
\end{figure*}

\begin{deluxetable*}{lccccc}\label{tab:z-score-ParamANN}
\tablecaption{$z$-scores and statistical consistencies for each of this paper's ANN model posteriors compared to ParamANN (reported values $\Omega_m=0.3029\pm0.1118$, $\Omega_\Lambda=0.6258\pm0.1689$).}
\tablehead{\colhead{Model/$k$} & \colhead{$z$ vs ParamANN $(\Omega_m)$} & \colhead{$z$ vs ParamANN $(\Omega_\Lambda)$} & \colhead{$z_{\rm sum}$} & \colhead{$D=\sqrt{z_m^2+z_\Lambda^2}$} & \colhead{Within $n\sigma$}}
\startdata
MLP ($k=0.5$) & 0.42 & 0.70 & 1.12 & 0.81 & $\le 1\sigma$ \\
MLP ($k=1.0$) & 0.36 & 0.66 & 1.02 & 0.75 & $\le 1\sigma$ \\
MLP ($k=2.0$) & 0.43 & 0.70 & 1.14 & 0.82 & $\le 1\sigma$ \\
CNN ($k=0.5$) & 0.36 & 0.66 & 1.01 & 0.75 & $\le 1\sigma$ \\
CNN ($k=1.0$) & 0.40 & 0.69 & 1.09 & 0.79 & $\le 1\sigma$ \\
CNN ($k=2.0$) & 0.55 & 0.79 & 1.34 & 0.96 & $\le 1\sigma$ \\
RNN (BiLSTM) ($k=0.5$) & 0.24 & 0.58 & 0.81 & 0.63 & $\le 1\sigma$ \\
RNN (BiLSTM) ($k=1.0$) & 0.31 & 0.62 & 0.94 & 0.69 & $\le 1\sigma$ \\
RNN (BiLSTM) ($k=2.0$) & 0.37 & 0.67 & 1.04 & 0.76 & $\le 1\sigma$ \\
CNN+RNN (BiLSTM) ($k=0.5$) & 0.29 & 0.61 & 0.90 & 0.67 & $\le 1\sigma$ \\
CNN+RNN (BiLSTM) ($k=1.0$) & 0.34 & 0.65 & 0.99 & 0.73 & $\le 1\sigma$ \\
CNN+RNN (BiLSTM) ($k=2.0$) & 0.42 & 0.70 & 1.13 & 0.81 & $\le 1\sigma$ \\
CNN+RNN (GRU) ($k=0.5$) & 0.27 & 0.60 & 0.87 & 0.66 & $\le 1\sigma$ \\
CNN+RNN (GRU) ($k=1.0$) & 0.32 & 0.63 & 0.96 & 0.71 & $\le 1\sigma$ \\
CNN+RNN (GRU) ($k=2.0$) & 0.37 & 0.67 & 1.03 & 0.76 & $\le 1\sigma$ \\
\enddata
\tablecomments{$z=|\mu_{\rm model}-\mu_{\rm ref}|/\sqrt{\sigma_{\rm model}^2+\sigma_{\rm ref}^2}$. $D=\sqrt{z(\Omega_m)^2+z(\Omega_\Lambda)^2}$ is the joint (2D) distance assuming independence. We compare $D^2$ to $\chi^2_{\nu=2}$: $1\sigma\Rightarrow D^2\le2.30$, $2\sigma\Rightarrow D^2\le6.18$, $3\sigma\Rightarrow D^2\le11.83$. In this case, all rows are well within $1\sigma$ for ParamANN.}
\end{deluxetable*}

\begin{deluxetable*}{lccccc}\label{tab:z-score-planck}
\tablecaption{$z$-scores and statistical consistencies for each of this paper's ANN model posteriors compared to Planck Collaboration (flat $\Lambda$CDM; $\Omega_m=0.315\pm0.007$, $\Omega_\Lambda=0.685\pm0.007$).}
\tablehead{\colhead{Model/$k$} & \colhead{$z$ vs Planck $(\Omega_m)$} & \colhead{$z$ vs Planck $(\Omega_\Lambda)$} & \colhead{$z_{\rm sum}$} & \colhead{$D=\sqrt{z_m^2+z_\Lambda^2}$} & \colhead{Within $n\sigma$}}
\startdata
MLP ($k=0.5$) & 3.37 & 3.17 & 6.54 & 4.63 & $>3\sigma$ \\
MLP ($k=1.0$) & 1.68 & 1.64 & 3.32 & 2.35 & $\le 2\sigma$ \\
MLP ($k=2.0$) & 1.07 & 1.07 & 2.14 & 1.51 & $\le 1\sigma$ \\
CNN ($k=0.5$) & 3.03 & 2.96 & 5.99 & 4.24 & $>3\sigma$ \\
CNN ($k=1.0$) & 3.88 & 3.92 & 7.80 & 5.52 & $>3\sigma$ \\
CNN ($k=2.0$) & 6.30 & 6.41 & 12.72 & 8.99 & $>3\sigma$ \\
RNN (BiLSTM) ($k=0.5$) & 4.37 & 4.51 & 8.87 & 6.28 & $>3\sigma$ \\
RNN (BiLSTM) ($k=1.0$) & 5.73 & 5.70 & 11.43 & 8.08 & $>3\sigma$ \\
RNN (BiLSTM) ($k=2.0$) & 6.73 & 6.95 & 13.68 & 9.67 & $>3\sigma$ \\
CNN+RNN (BiLSTM) ($k=0.5$) & 5.01 & 4.98 & 9.99 & 7.06 & $>3\sigma$ \\
CNN+RNN (BiLSTM) ($k=1.0$) & 5.44 & 5.41 & 10.85 & 7.67 & $>3\sigma$ \\
CNN+RNN (BiLSTM) ($k=2.0$) & 6.37 & 6.39 & 12.77 & 9.02 & $>3\sigma$ \\
CNN+RNN (GRU) ($k=0.5$) & 4.76 & 4.66 & 9.42 & 6.66 & $>3\sigma$ \\
CNN+RNN (GRU) ($k=1.0$) & 5.50 & 5.35 & 10.85 & 7.67 & $>3\sigma$ \\
CNN+RNN (GRU) ($k=2.0$) & 5.27 & 5.58 & 10.86 & 7.68 & $>3\sigma$ \\
\enddata
\tablecomments{$z=|\mu_{\rm model}-\mu_{\rm ref}|/\sqrt{\sigma_{\rm model}^2+\sigma_{\rm ref}^2}$. $D=\sqrt{z(\Omega_m)^2+z(\Omega_\Lambda)^2}$ is the joint (2D) distance assuming independence. We compare $D^2$ to $\chi^2_{\nu=2}$: $1\sigma\Rightarrow D^2\le2.30$, $2\sigma\Rightarrow D^2\le6.18$, $3\sigma\Rightarrow D^2\le11.83$.}
\end{deluxetable*}

\section{Conclusion}\label{sec:conclusion}

We investigated five ANN architectures - MLP, CNN, BiLSTM RNN, and two CNN+RNN hybrids (BiLSTM/GRU) - for inferring $(\Omega_{m,0}, \Omega_{\Lambda,0})$ from synthetic $H(z)$ under flat $\Lambda$CDM with fixed $H_0$. We constructed a data-driven noise model using KDE on the relative uncertainties of 47 low-$z$ observations and generated training sets across three SNR regimes ($k=0.5,1.0,2.0$). Model performance was assessed by a balanced suite of distributional and pointwise metrics (two-sample KS, $W_1$, $\chi^2_{\rm obs}$, and GLS $\chi^2$), aggregated via a ranking scheme to avoid overemphasizing correlated statistics.

Across metrics, higher-SNR training (\mbox{$k=0.5$}) consistently improved agreement with observed $H(z)$ distributions. The BiLSTM RNN achieved the best overall score at high SNR, while the CNN+GRU hybrid was most robust at low SNR, reflecting complementary strengths in local pattern extraction and sequence dependence. Notably, while all architectures remained statistically consistent with ParamANN \citep{Pal2024} within $1\sigma$, most configurations exhibited $>3\sigma$ tension with \citet{Planck2020} posteriors under our modeling choices. This highlights a key gap between synthetic low-$z$ training plus simplified assumptions and the multi-probe and high-precision constraints implicit in \citet{Planck2020}.

Therefore, these limitations point to clear next steps: (i) make the model more comprehensive by broadening priors and testing more free parameters such as $H_0$ (to analyze the Hubble tension problem) and $\Omega_k$ (to evaluate in a \text{wCDM} universe model); (ii) stress-testing the KDE noise law versus redshift-dependent systematics; (iii) report coverage and calibration diagnostics (e.g., PIT histograms, empirical credible-interval coverage) beyond pointwise and distributional fits; and (iv) extend the domain to $z>1$ for wider applicability; With these additions, ANN-based cosmological inference can become more accurate, applicable to more astrophysical scenarios, and more comparable to classical MCMC pipelines, while retaining its efficiency and flexibility.


\clearpage

\bibliography{sample701}
\bibliographystyle{aasjournalv7}

\end{document}